# ADOPTION FACTORS OF ENABLING I4.0 TECHNOLOGIES AND BENEFITS IN THE SUPPLY CHAIN

José Carlos Franceli and Silvia Novaes Zilber Turri

Universidade Federal do ABC, São Paulo, Brazil

## ABSTRACT

*Industry 4.0 technologies represent a new paradigm of integration of cyber-physical systems, information and communication solutions, with applications in many different domains. This topic has to date been vaguely explored in the realm of the social sciences; hence this study attempts to bridge this gap by investigating the challenges of innovation adoption, based on I4.0 technologies, more specifically the factors affecting adoption decisions. This paper, based on previous adoption literature, aims to identify the barriers and benefits generated in the Supply Chain. Given the nature and novelty of the technology, whose adoption is the primary theme of this study, a systematic literature review was developed. The results present a framework that connects adoption factors, enabling technologies of I4.0, and benefits to the Supply Chain. The Model can be easily adapted to serve as a tool in the evaluation and selection of technological innovations to be adopted.*

## 1. INTRODUCTION

The new Fourth Industrial Revolution, also known as Industry 4.0 (I4.0), is disrupting the way companies do business all around the world, where traditional manufacturing methods and production are undergoing digital transformation. The concept of Industry 4.0 was initially introduced in Germany in 2011 [19],referring to the integration of physical objects, human actors, intelligent machines, production lines and processes across organizational boundaries, with the aim of creating a system where processes are integrated and information is shared in real time [14]. The emergence of I4.0 is deeply rooted in the Third Industrial Revolution, characterized by rapid developments in information technology (IT), electronics and digitalization, where Advanced Manufacturing Technologies (AMTs) are at their core. AMTs can be described as computer-assisted technologies used to control and monitor manufacturing activities, providing greater flexibility, shorter production cycles, faster responses to changing market demands, better precision, and control of production processes [7]. IoT is revolutionizing the manufacturing industry and the consumption of goods; products that once were exclusively composed of mechanical and electrical parts are now becoming complex systems combining hardware, data storage, sensors, microprocessors, software, and connectivity in various formats. This is how the Internet of Things (IoT) is becoming widespread. Smart and connected products, enabled by major improvements in the processing of device miniaturization and wireless connectivity, are currently unleashing a new era of competition [25]. However, the wide spectrum of applications for the same technology can be, at the same time, an incentive and an obstacle to its adoption [5].

The adoption of innovation is a complex process especially when the innovation or technology in question is incipient [26] and several authors talk about the benefits it can generate when overcoming technological challenges. However, none of these authors discussed the challenges of adoption in different environments and market segments where solutions may be applied. The lack of a model for studying the adoption of this technology led us to the elaboration of a structure that contemplates the several dimensions related to the adoption decision. Therefore, this study, through a Systematic Literature Review (SLR), aims to identify the factors that lead to the adoption of Enabling Technologies of Industry4.0 (I4.0 technologies) and the benefits that they bring to the Supply Chain (SC). The following sections comprise the research method, followed by SLR results, the proposition of a conceptual framework and final considerations.

## 2. METHOD

### 2.1. Research Planning and Keyword Identification

This article adopts the methodology proposed in the stages of planning, execution and presentation of results obtained through this SLR [31]. The keywords selected represent the terms available in the literature of Industry 4.0 Technology Adoption in SC. The search for relevant articles was done through a keyword-combination process to identify the main study themes. The initial search utilized the keyword combination "Digital AND Transformation AND Adoption" to look for articles in the scope of our study objective. Subsequently, to expand the range of studies related to I4.0 technology adoption, a second search was carried out using the keyword combination "I4.0 AND Adoption", resulting in a group of relevant articles primarily focused on the topic of adoption in the field of I4.0 technologies. Finally, a third search was undertaken using the keyword combination "IoT AND Supply AND Chain AND Adoption" resulting in a broad search, combining keywords that brought us the results outlined in the study objectives.

### 2.2. Conducting the Search for Articles

To systematically evaluate the presented theme in the literature, this analytical review process requires relevant articles on the topic [22]. Following this process, it is possible to identify the main relevant publications, trends on the topics being discussed and researched, as well as to evidence the gaps present in the literature [30]. This session presents the research protocol used as part of the strategy to identify relevant studies and the criteria for inclusion and exclusion of previously selected documents [17]. Initially, from reading the "abstracts" of the selected articles, it was found that those who had "adoption" as the main theme resulted in a group of relevant articles for the study. We applied a second selection process on this group of articles to identify the most relevant articles according to content (adoption barriers and benefits), results (adoption framework) and type of organizations (manufacturing companies, transport services and retail).

## 3. RESULTS

### 3.1. Results Presentation

To have a detailed presentation of the study results, this section is structured in the following manner:

- Descriptive Statistics Analysis
- Content Analysis
- Summary of Results

## 3.2. Descriptive Statistics Analysis

Twenty-three (23) relevant articles were selected for this study after an analysis of ninety-three (93) abstracts from the articles found through the search using the three-keyword combination procedure. The 226 citations found in the 23 articles represent 41% of the total citation quote (556) between 2018 to 2020 (Figure 1). The evolution of citations for articles with "adoption" as a subject for the same period is show in Figure 2.

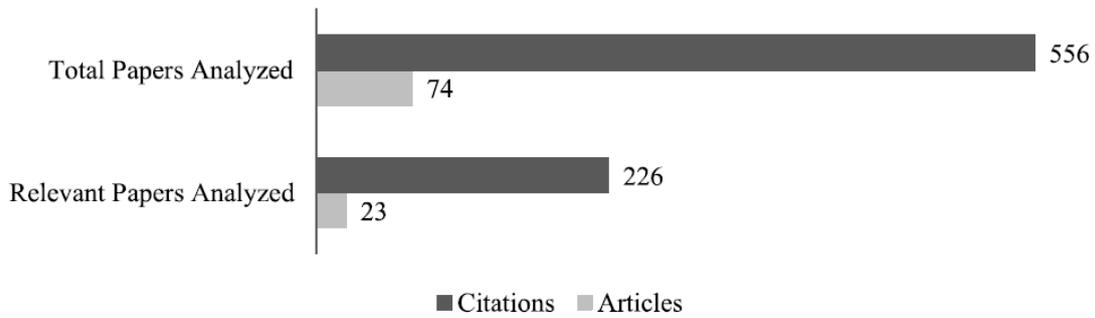

Figure 1. Relevant papers according to number of citations

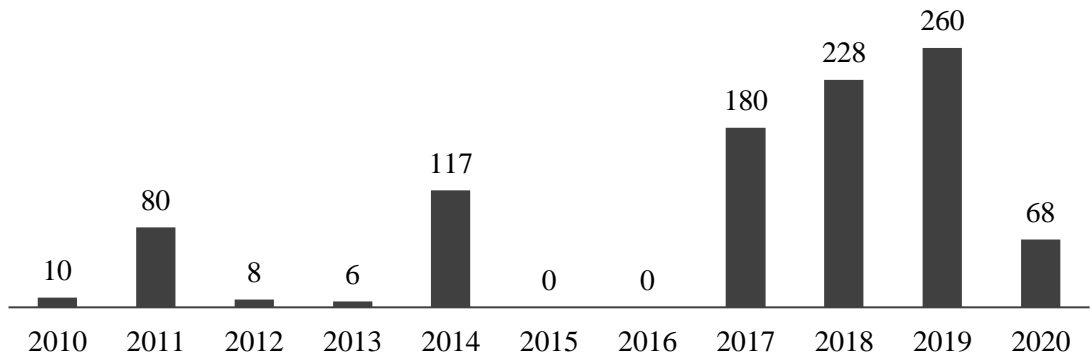

Figure 2. Number of citations in articles with "Adoption" as a theme over time

After a ranking process, the last prioritization selection was conducted to identify the ten (10) most relevant articles to the theme of this paper (Table 1) following the criteria below:

a. Articles that present adoption frameworks.
b. Articles that present factors and benefits of adoption.
c. Articles that reference companies in the manufacturing, logistics services and retail sectors.

Table 1. Selected articles and criteria

| Criteria | Number of Articles |
|---|---|
| a.   Articles that present adoption frameworks. | 4 |
| b.   Articles that present factors and benefits of adoption. | 10 |
| c.   Articles that reference companies in the manufacturing, logistics services and retail sectors. | 10 |

Figure3. outlines the complete research and selection methodology of articles used in this study.

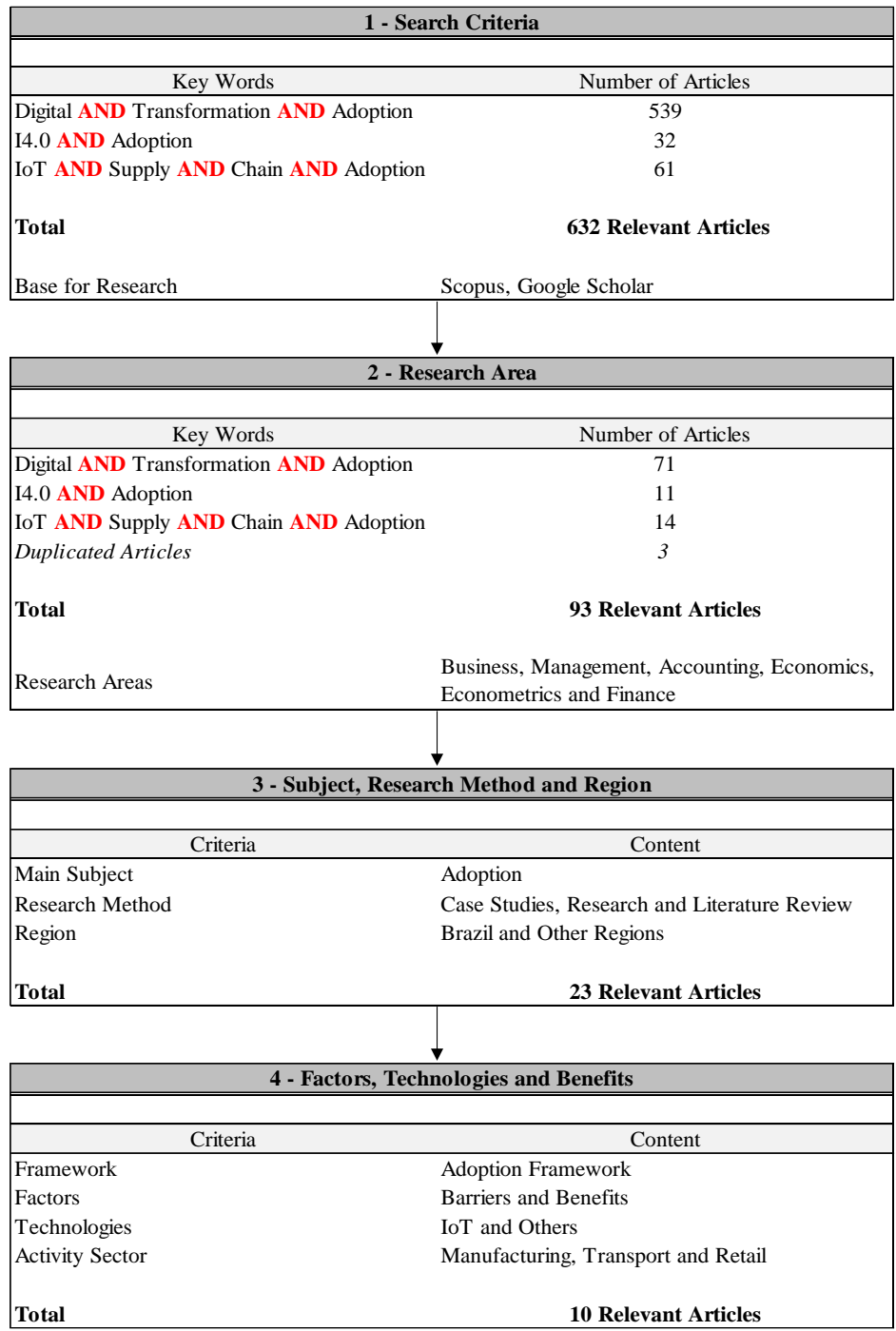

Figure 3. Research methodology

## 3.2.1.   List of Articles According to Keyword Combinations

The list below presents the 93 articles found in the research process.

| Article | Title | Author | Source |
|---|---|---|---|
| 1 | Moving towards digitalization: a multiple case study in manufacturing | Zangiacomi, A., Pessot, E., Fornasiero, R., Bertetti, M., Sacco, M. | Production Planning and Control 31(2-3), pp. 143-157 |
| 2 | Organizational learning paths based upon industry 4.0 adoption: An empirical study with Brazilian manufacturers | Tortorella, G.L., Cawley Vergara, A.M., Garza-Reyes, J.A., Sawhney, R. | International Journal of Production Economics 219, pp. 284-294 |
| 3 | Bringing it all back home? Backshoring of manufacturing activities and the adoption of Industry 4.0 technologies | Dachs, B., Kinkel, S., Jäger, A. | Journal of World Business 54(6),101017 |
| 4 | Providing industry 4.0 technologies: The case of a production technology cluster | Dalmarco, G., Ramalho, F.R., Barros, A.C., Soares, A.L. | Journal of High Technology Management Research 30(2),100355 |
| 5 | The adoption of Industry 4.0 technologies in SMEs: results of an international study | Agostini, L., Nosella, A. | 58(4), pp. 625-643 Management Decision |
| 6 | Organizational and managerial challenges in the path toward Industry 4.0 | Agostini, L., Filippini, R. | European Journal of Innovation Management 22(3), pp. 406-421 |
| 7 | The impacts of Industry 4.0: a descriptive survey in the Italian manufacturing sector | Zheng, T., Ardolino, M., Bacchetti, A., Perona, M., Zanardini, M. | Journal of Manufacturing Technology Management   Article in Press |
| 8 | A comparison on Industry 4.0 and Lean Production between manufacturers from emerging and developed economies | Tortorella, G.L., Rossini, M., Costa, F., Portioli Staudacher, A., Sawhney, R. | Total Quality Management and Business Excellence   Article in Press |
| 9 | Fashion 4.0. Innovating fashion industry through digital transformation | Bertola, P., Teunissen, J. | Research Journal of Textile and Apparel 22(4), pp. 352-369 |
| 10 | New Model for Identifying Critical Success Factors Influencing BIM Adoption from Precast Concrete Manufacturers' View | Phang, T.C.H., Chen, C., Tiong, R.L.K. | Journal of Construction Engineering and Management 146(4),04020014 |
| 11 | Exploring the growth challenge of mobile payment platforms: A business model perspective | Jocevski, M., Ghezzi, A., Arvidsson, N. | Electronic Commerce Research and Applications |
| 13 | Digital transformation in supply chain, challenges and opportunities in SMEs: a case study of Al-Rumman Pharma | Faridi, M.R., Malik, A. | Emerald Emerging Markets Case Studies |
| 14 | Digital systems and new challenges of financial management – fintech, XBRL, blockchain and cryptocurrencies | Mosteanu, N.R., Faccia, A. | Quality - Access to Success |
| 15 | How enterprises adopt agile forms of organizational design: A multiple-case study | Gerster, D., Dremel, C., Brenner, W., Kelker, P. | Data Base for Advances in Information Systems |
| 16 | Analysis of barriers in implementation of digital transformation of supply chain using interpretive structural modelling approach | Agrawal, P., Narain, R., Ullah, I. | Journal of Modelling in Management |
| 17 | Fostering digital transformation of SMEs: a four levels approach | Garzoni, A., De Turi, I., Secundo, G., Del Vecchio, P. | Management Decision |
| 18 | Digital transformation priorities of India's discrete manufacturing SMEs – a conceptual study in perspective of Industry 4.1 | Dutta, G., Kumar, R., Sindhwani, R., Singh, R.K. | Competitiveness Review |
| 19 | Digital initiatives for access and quality in higher education: An overview | Ahmad, S. | Prabandhan: Indian Journal of Management |
| 20 | PERCEPTION or CAPABILITIES? AN EMPIRICAL INVESTIGATION of the FACTORS INFLUENCING the ADOPTION of SOCIAL MEDIA and PUBLIC CLOUD in German SMEs | Hassan, S.S., Reuter, C., Bzhalava, L. | International Journal of Innovation Management |
| 21 | Organizational learning paths based upon industry 4.0 adoption: An empirical study with Brazilian manufacturers | Tortorella, G.L., Cawley Vergara, A.M., Garza-Reyes, J.A., Sawhney, R. | International Journal of Production Economics |
| 22 | Digitalization of construction organisations–a case for digital partnering | Aghimien, D., Aigbavboa, C., Oke, A., Thwala, W., Moripe, P. | International Journal of Construction Management |
| 23 | What matters in implementing the factory of the future: Insights from a survey in European manufacturing regions | Pessot, E., Zangiacomi, A., Battistella, C., (...), Sala, A., Sacco, M. | Journal of Manufacturing Technology Management |
| 24 | Digitalization of world trade: Scope, forms, implications | Strelets, I.A., Chebanov, S.V. | World Economy and International Relations |
| 25 | The impact of Industry 4.0 implementation on supply chains | Ghadge, A., Er Kara, M., Moradlou, H., Goswami, M. | Journal of Manufacturing Technology Management |
| 26 | Big data and HR analytics in the digital era | Dahlbom, P., Siikanen, N., Sajasalo, P., Jarvenpää, M. | Baltic Journal of Management |
| 27 | Conceptualizing digital transformation in SMEs: an ecosystemic perspective | Pelletier, C., Cloutier, L.M. | Journal of Small Business and Enterprise Development |
| 28 | Tortoise, not the hare: Digital transformation of supply chain business processes | Hartley, J.L., Sawaya, W.J. | Business Horizons |
| 29 | Readiness factors for information technology adoption in SMEs: testing an exploratory model in an Indian context | Nair, J., Chellasamy, A., Singh, B.N.B. | Journal of Asia Business Studies |
| 30 | Digital passengers: A great divide or emerging opportunity? | Mayer, C. | Journal of Airport Management |
| 31 | A systematic literature review of big data adoption in internationalization | Dam, N.A.K., Le Dinh, T., Menvielle, W. | Journal of Marketing Analytics |
| 32 | The change of pediatric surgery practice due to the emergence of connected health technologies | Niemelä, R., Pikkarainen, M., Ervasti, M., Reponen, J. | Technological Forecasting and Social Change |
| 33 | To be or not to be digital, that is the question: Firm innovation and performance | Ferreira, J.J.M., Fernandes, C.I., Ferreira, F.A.F. | Journal of Business Research |
| 34 | Industry 4.0 and capability development in manufacturing subsidiaries | Szalavetz, A. | Technological Forecasting and Social Change |
| 35 | Do German Works Councils Counter or Foster the Implementation of Digital Technologies? : First Evidence from the IAB-Establishment Panel | Genz, S., Bellmann, L., Matthes, B. | Jahrbucher fur Nationalokonomie und Statistik |
| 36 | Technology adoption for the integration of online–offline purchasing: Omnichannel strategies in the retail environment | Lobo, F., VASCONCELLOS, E., & Guedes, L. V | International Journal of Retail and Distribution Management |
| 37 | Readiness of upscale and luxury-branded hotels for digital transformation | Lam, C., Law, R. | International Journal of Hospitality Management |
| 38 | Blockchain technology and its relationships to sustainable supply chain management | Saberi, S., Kouhizadeh, M., Sarkis, J., Shen, L. | International Journal of Production Research |
| 39 | Disruptions of account planning in the digital age | Zimand Sheiner, D., Earon, A. | Marketing Intelligence and Planning |
| 40 | Industry 4.0 technologies: Implementation patterns in manufacturing companies | Frank, A.G., Dalenogare, L.S., Ayala, N.F. | International Journal of Production Economics |
| 41 | Exploring the digital innovation process: The role of functionality for the adoption of innovation management software by innovation managers | Huesig, S., Endres, H. | European Journal of Innovation Management |
| 42 | Opportunities and challenges in the e-commerce of the food sector | Vargas, V.M., Budz, S. | Quality - Access to Success |
| 43 | Fintechs: A literature review and research agenda | Milian, E.Z., Spinola, M.D.M., Carvalho, M.M.D. | Electronic Commerce Research and Applications |
| 44 | Transformation of accounting through digital standardisation: Tracing the construction of the IFRS Taxonomy | Troshani, I., Locke, J., Rowbottom, N. | Accounting, Auditing and Accountability Journal |
| 45 | The effective factors of cloud computing adoption success in organization | Yoo, S.-K., Kim, B.-Y. | Journal of Asian Finance, Economics and Business |
| 46 | The impact of digital transformation on formal and informal organizational structures of large architecture and engineering firms | Bonanomi, M.M., Hall, D.M., Staub-French, S., Tucker, A., Talamo, C.M.L. | Engineering, Construction and Architectural Management |
| 47 | Digital transformation in the Spanish agri-food cooperative sector: Situation and prospects | [La transformación digital en el sector cooperativo agroalimentario español: Situación y perspectivas] | Vázquez, J.J., Cebolla, M.P.C., Ramos, F.S. | CIRIEC-Espana Revista de Economía Publica, Social y Cooperativa |

| # | Title | Authors | Journal |
|---|-------|---------|---------|
| 48 | Bridging the gender digital gap | Mariscal, J., Mayne, G., Aneja, U., Sorgner, A. | Economics |
| 49 | Disruptive technology adoption, particularities of clustered firms | Molina-Morales, F.X., Martínez-Cháfer, L., Valiente-Bordanova, D. | Entrepreneurship and Regional Development |
| 50 | Bitcoin distribution in the age of digital transformation: Dual-path approach | Lee, W.-J., Hong, S.-T., Min, T. | Journal of Distribution Science |
| 51 | Determinants of information and digital technology implementation for smart manufacturing | Ghobakhloo, M. | International Journal of Production Research |
| 52 | Embracing artificial intelligence and digital personnel to create high-performance jobs in the cyber economy | Lobova, S.V., Bogoviz, A.V. | Contributions to Economics |
| 53 | The impacts of Industry 4.0: a descriptive survey in the Italian manufacturing sector | Zheng, T., Ardolino, M., Bacchetti, A., Perona, M., Zanardini, M. | Journal of Manufacturing Technology Management |
| 54 | Blockchain: A paradigm shift in business practices | Kizildag, M., Dogru, T., Zhang, T., (...), Ozturk, A.B., Ozdemir, O. | International Journal of Contemporary Hospitality Management |
| 55 | Extremes of acceptance: employee attitudes toward artificial intelligence | Lichtenthaler, U. | Journal of Business Strategy |
| 56 | Fashion 4.0. Innovating fashion industry through digital transformation | Bertola, P., Teunissen, J. | Research Journal of Textile and Apparel |
| 57 | The impacts of digital transformation on the labour market: Substitution potentials of occupations in Germany | Dengler, K., Matthes, B. | Technological Forecasting and Social Change |
| 58 | It consumerization and the transformation of it governance | Gregory, R.W., Kaganer, E., Henfridsson, O., Ruch, T.J. | MIS Quarterly: Management Information Systems |
| 59 | Loosely Coupled Systems of Innovation: Aligning BIM Adoption with Implementation in Dutch Construction | Papadonikolaki, E. | Journal of Management in Engineering |
| 60 | New technologies and the transformation of work and skills: a study of computerisation and automation of Australian container terminals | Gekara, V.O., Thanh Nguyen, V.-X. | New Technology, Work and Employment |
| 61 | Active seniors perceived value within digital museum transformation | Traboulsi, C., Frau, M., Cabiddu, F. | TQM Journal |
| 62 | Insights on the adoption of social media marketing in B2B services | Buratti, N., Parola, F., Satta, G. | TQM Journal |
| 63 | Digital technologies and the modernization of public administration | Todorut, A.V., Tselentis, V. | Quality - Access to Success |
| 64 | Digital transformation at carestream health | Smith, H.A., Watson, R.T. | MIS Quarterly Executive |
| 65 | IT Governance Mechanisms and Contingency Factors: Towards an Adaptive IT Governance Model | Levstek, A., Hovelja, T., Pucihar, A. | Organizacija |
| 66 | The impact of digital technology on relationships in a business network | Pagani, M., Pardo, C. | Industrial Marketing Management |
| 67 | Making Sense of Africa's Emerging Digital Transformation and its Many Futures | Ndemo, B., Weiss, T. | Africa Journal of Management |
| 68 | Work 4.0 — Digitalisation and its Impact on the Working Place | [Arbeiten 4.0 — Folgen der Digitalisierung für die Arbeitswelt] | Klammer, U., Steffes, S., Maier, M.F., (...), Bellmann, L., Hirsch-Kreinsen, H. | Wirtschaftsdienst |
| 69 | How transformational leadership facilitates e-business adoption | Akoi-Simo, L., Verdu-Jover, A.J., Gomez-Gras, J.-M. | Industrial Management and Data Systems |
| 70 | Tackling the digitalization challenge: How to benefit from digitalization in practice | Parviainen, P., Tihinen, M., Kääriäinen, J., Teppola, S. | International Journal of Information Systems and Project Management |
| 71 | Are millennials transforming global tourism? Challenges for destinations and companies | Veiga, C., Santos, M.C., Águas, P., Santos, J.A.C. | Worldwide Hospitality and Tourism Themes |
| 72 | Introducing data driven practices into sales environments: examining the impact of data visualisation on user engagement and sales results | Magee, B., Sammon, D., Nagle, T., O'Raghallaigh, P. | Journal of Decision Systems |
| 73 | Evaluating data driven practices in sales environments: user engagement and sales results | Magee, B. | Journal of Decision Systems |
| 74 | The effects of rfid applied in ground handling system on aircraft turnaround time: A simulation based analysis | Khumboon, R., Isaradech, B. | Academy of Strategic Management Journal |
| 75 | Exploring the relevancy of Massive Open Online Courses (MOOCs): A Caribbean university approach | Dyer, R.A.D. | Information Resources Management Journal |
| 76 | Paradoxical digital worlds | Munar, A.M. | Tourism Social Science Series |
| 77 | The role of ICTs in conflict transformation in Egypt | Richardson, J.W., Brantmeier, E.J. | Education, Business and Society: Contemporary Middle Eastern Issues |
| 78 | Analysis of emerging technology adoption for the digital content market | Jin, B.-H., Li, Y.-M. | Information Technology and Management |
| 79 | Student attitudes and behaviors towards digital textbooks | Weisberg, M. | Publishing Research Quarterly |
| 80 | Fundamentals of H.324 desktop videoconferencing | Herman, Mort | Electronic Design |
| 81 | Technology embracing by 3pl service providers in india: Tuticorin port trust – a case study | Senthil, M., Ruthramathi, R., Gayathri, N. | International Journal of Scientific and Technology Research 9(3), pp. 138-144 |
| 82 | Managing the food supply chain in the age of digitalisation: a conceptual approach in the fisheries sector | Coronado Mondragon, A.E., Coronado Mondragon, C.E., Coronado, E.S. | Production Planning and Control |
| 83 | Modeling the internet of things adoption barriers in food retail supply chains | Kamble, S.S., Gunasekaran, A., Parekh, H., Joshi, S. | Journal of Retailing and Consumer Services 48, pp. 154-168 |
| 84 | The Internet of Things (IoT) in retail: Bridging supply and demand | Raman, S., Patwa, N., Niranjan, I., (...), Moorthy, K., Mehta, A. | Journal of Retailing and Consumer Services 48, pp. 154-169 |
| 85 | Adoption of Internet of Things in India: A test of competing models using a structured equation modeling approach | Mital, M., Chang, V., Choudhary, P., Papa, A., Pani, A.K. | ITechnological Forecasting and Social Change 136, pp. 339-346 |
| 85 | Impact of big data on supply chain management | Raman, S., Patwa, N., Niranjan, I., (...), Moorthy, K., Mehta, A. | International Journal of Logistics Research and Applications 21(6), pp. 579-596 |
| 86 | Factors influencing the adoption of the internet of things in supply chains | Yan, B., Jin, Z., Liu, L., Liu, S. | Journal of Evolutionary Economics |
| 87 | Real-Time business data acquisition: How frequent is frequent enough? | Townsend, M., Le Quoc, T., Kapoor, G., (...), Zhou, W., Piramuthu, S. | Information and Management 55(4), pp. 422-429 |
| 88 | The effect of "Internet of Things" on supply chain integration and performance: An organisational capability perspective | Tsang, Y.P., Choy, K.L., Wu, C.H., (...), Lam, H.Y., Koo, P.S. | Australasian Journal of Information Systems 22 |
| 89 | Internet of Things (IoT) in high-risk Environment, Health and Safety (EHS) industries: A comprehensive review | Thibaud, M., Chi, H., Zhou, W., Piramuthu, S. | Decision Support Systems 108, pp. 79-95 |
| 90 | An IoT-based cargo monitoring system for enhancing operational effectiveness under a cold chain environment | | International Journal of Engineering Business Management |
| 91 | Examining potential benefits and challenges associated with the Internet of Things integration in supply chains | Haddad, A., DeSouza, A., Khare, A., Lee, H. | Journal of Manufacturing Technology Management |
| 92 | Impact of RFID technology on supply chain decisions with inventory inaccuracies | Fan, T., Tao, F., Deng, S., Li, S. | International Journal of Production Economics 159, pp. 117-125 |
| 93 | Integrated billing mechanisms in the Internet of Things to support information sharing and enable new business opportunities | Uckelmann, D., Harrison, M. | International Journal of RF Technologies: Research and Applications 2(2), pp. 73-90 |

The 10 articles below form the basis of the SLR selected by prioritization criteria:

| Article Number | Title | Author | Source |
|---|---|---|---|
| 1 | The adoption of Industry 4.0 technologies in SMEs: results of an international study | Agostini, L., Nosella, A. | Management Decision |
| 2 | Organizational and managerial challenges in the path toward Industry 4.0 | Agostini, L., Filippini, R. | European Journal of Innovation Management 22(3), pp. 406-421 |
| 3 | Perception or Capabilities? An Empirical Investigation of the Factors Influencing the Adoption of Social Media and Public Cloud in German SMEs | Hassan, S.S., Reuter, C., Bzhalava, L. | International Journal of Innovation Management |
| 4 | The impact of Industry 4.0 implementation on supply chains | Ghadge, A., Er Kara, M., Moradlou, H., Goswami, M. | Journal of Manufacturing Technology Management |
| 5 | To be or not to be digital, that is the question: Firm innovation and performance | Ferreira, J.J.M., Fernandes, C.I., Ferreira, F.A.F. | Journal of Business Research |
| 6 | Technology adoption: Factors influencing the adoption decision of the internet of things in a business environment | LOBO, F., VASCONCELLOS, E., & GUEDES, L. | International Association for Management of Technology |
| 7 | Disruptive technology adoption, particularities of clustered firms | Molina-Morales, F.X., Martínez-Cháfer, L., Valiente-Bordanova, D. | Entrepreneurship and Regional Development |
| 8 | Modeling the internet of things adoption barriers in food retail supply chains | Kamble, S.S., Gunasekaran, A., Parekh, H., Joshi, S. | International Journal of Production Research |
| 9 | Adoption of Internet of Things in India: A test of competing models using a structured equation modeling approach | Mital , M., Chang, V., Choudhary , P., Papa, A., Pani , AK | ITechnological Forecasting and Social Change 136, pp. 339-346 |
| 10 | Factors influencing the adoption of the internet of things in supply chains | Yan, B., Jin, Z., Liu, L., Liu, S. | Journal of Evolutionary Economics |

### 3.2.2. Statistics of Adoption Factors, Technologies and Benefits

As shown in Table 2, the results from the SLR found nineteen (19) factors impacting the adoption process.

Table 2. Adoption factor per analyzed study (authors)

| ADOPTION FACTORS / AUTHORS | Hassan, S.S., Reuter, C., Bzhalava, L. | Kamble, S.S., Gunasekaran, A., Parekh, H., Joshi, S. | Agostini, L., Nosella, A. | Mital, M., Chang, V., Choudhary, P., Papa, A., Pani, A.K. | Ghadge, A., Er Kara, M., Moradlou, H., Goswami, M. | Agostini, L., Filippini, R. | Ferreira, J.J.M., Fernandes, C.I., Ferreira, F.A.F. | Molina-Morales, F.X., Martínez-Cháfer, L., Valiente-Bordanova, D. | Yan, B., Jin, Z., Liu, L., Liu, S. | Lobo, F., VASCONCELLOS, E., & Guedes, L. V |
|---|---|---|---|---|---|---|---|---|---|---|
| Perception of Importance of Use | x | | | x | | | | | | |
| Perception of Easy of Use | | | | x | | | | | | |
| Encouragement of Use by Other Users | | | | x | | | | | | |
| Adopter's Profile | | | | | | | | x | | |
| Employee Skills | x | x | x | | | x | | | | |
| Innovation and Digital Culture | x | | | | x | | | | x | |
| Security and Privacy | x | x | | | | | | | | |
| Acquisition and Operating Costs | x | x | | | x | | | | x | |
| Social Capital | | | x | | x | x | | | | |
| Management Support | | | | x | x | x | | | | |
| Absorptive Capacity | | | x | | | | | | x | x |
| Sector Activity | x | | | | | | x | | | |
| Market Demand | | x | | | | | | | | |
| Government Policies and Regulations | | x | | | x | | x | x | | |
| Standards and Validations | | x | | | x | | | | | x |
| IT Infrastructure | | x | | | x | | | | | |
| Systems Architecture, Integration and Compatibility | | x | | | x | | | | | x |
| Research and Development | | | | | x | | | | | |
| Data Management Quality | | | | | x | | | | | |

Following a structural adaptation from the adoption framework by Lobo *et al* [18], the 19 factors were grouped into four (4) dimensional clusters according to their nature: "Human", "Organizational", "Political-Market" and "Technological". Figure 4shows the 4 clusters and their respective adoption factors according to the frequency in which they appear in the studies.

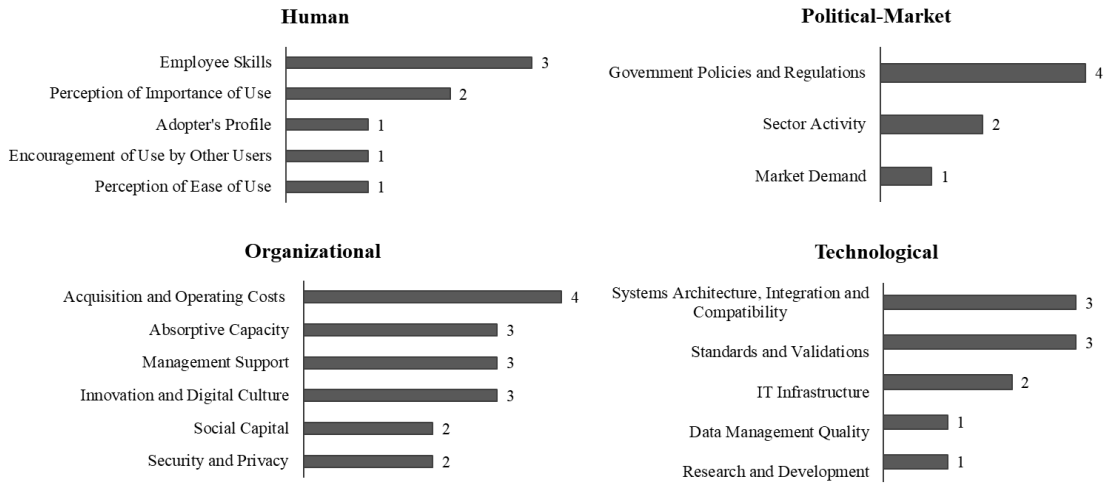

Figure 4. Frequency of factors per dimension

## 3.3. Content Analysis

In this section of the study, a content analysis is undertaken on the main adoption factors and benefits of I4.0 technology adoption based on the results drawn from the 10 articles of the SLR.

### 3.3.1.   Human Factors

#### 3.3.1.1. Perception of Importance of Technology Use

The "Perception of Importance of Use" acts as a high-relevance factor in technology adoption [28], as decision-makers tend to adopt innovations that are deemed useful and consistent with the organizational configurations processes [29].

#### 3.3.1.2. Perception of Ease of Technology Use

The Technology Acceptance Model (TAM) [7] shows that individuals' Perceived Usefulness (PU) of the technology and its Perceived Ease of Use (PEOU) impact the Behavioral Intention (BI) to use the technology. PEOU proved to be even more impactful than PU in the technology adoption process [31].

#### 3.3.1.3. Encouragement of Technology Use by Other Users

The studies using the TRA (Theory of Reasoned Action) Model and the "Physical and Behavioral Concept" [3], identified that the encouragement of technology use by other users, who also share the same perception, impacts the intended adoption and use of technologies [31].

### 3.3.1.4. Adopter's Profile

Factors related to personal characteristics of entrepreneurs, *i.e.* factors of non-economic nature, explain the adoption behavior of companies regarding digital processes [11]. More experienced entrepreneurs and managers have a lower propensity to adopt digital technologies; while female managers, graduates and university students are more prone to adopting these technologies [10]. Therefore, the "Adopter's Profile" plays an important role in the adoption of I4.0 technologies.

### 3.3.1.5. Employee Skills

In the study conducted by Kamble *et al* [16], 12 categories were identified to impact I4.0 technologies adoption, more specifically IoT. Among these 12 categories, one of the most relevant was human skills, which can influence the entire structure of the adoption process [16]. IoT systems require highly trained professionals to develop and implement practical applications [28]. Qualified employees in innovation processes, who have advanced innovation skills and understand the use of lean methods and Information and Communication Technologies (ICT), pave the way for the adoption of I4.0 technologies [1]. Moreover, training and professional development are essential in the initial stages of the transition to digitalization [1].

### 3.3.2.     Organizational Factors

### 3.3.2.1. Innovation and Digital Culture

Digital technology is changing business practice and organizational culture around the world. The convergence of modern digital technology, *e.g.* I4.0 technologies, is widely discussed as the next source of innovation and productivity in organizations. These are enabled by organizations' capabilities and innovation resources [29]. The change in business ecosystems due to innovation profoundly influences operational models and structures, and management strategies to adapt and explore new challenges in this evolving landscape [12].

Early adopters of I4.0 technologies, who are highly oriented toward implementing early-stage technologies, that work in clusters, can also multiply the likelihood of adoption by 473.7 per cent. The grouping process in clusters generates tacit knowledge and exchange of high-quality information, hence promoting innovation [23].

### 3.3.2.2. Security and Privacy

Given the existing threats in the virtual world, "Security and Privacy" are two of the most critical factors in the decision to adopt I4.0 technologies. Security is in danger in a network-based system due to threats of substitution by false data, access to confidential data and many other unauthorized intrusions that can paralyze networks [24]. Generally, companies internalize their security and privacy specifications, despite there being a lack of qualified competences and diversified capabilities related to the implementation of technologies beyond the scope of organizations' operations [29]. Moreover, the uncertainties associated with security and privacy lead decision-makers in small and medium enterprises (SMEs) to be reluctant to adopt I4.0 technologies or simply to ignore the potential benefit of these technologies for their businesses [8].

### 3.3.2.3. Acquisition and Operating Costs

"Acquisition Costs" and un foreseen "Operating Costs" of I4.0 technologies are some of the most relevant factors in the adoption decision, mainly to SMEs, and in many cases can lead to a rejection of the technology [21]. Operating costs should be considered as a high priority despite the upfront uncertainty of the magnitude of these costs, *i.e.* generally related to energy consumption, operations, and a long payback period.

There is a critical risk related to the financial loss and irreversibility of investment [9]. Additionally, the implementation of IoT solutions requires advanced techniques and infrastructure support [16]. Therefore, SMEs due to limited financial resources are generally cautious with hidden costs and un accounted expenses, especially in the absence or shortage of skilled labor to operate new technologies [29].

### 3.3.2.4. Social Capital

The successful adoption of I4.0 technologies often requires reliable cooperation, not only within the boundaries of the organization but also with external stakeholders [12]. Internal social capital represents the shared beliefs and values that allow individuals within an organization to work toward a common purpose, whereas external social capital involves the shared principles among external stakeholders along the SC. The impact of "Social Capital" in the adoption process is structured in 2 phases: digitalization of the business and transformation of the business network [17]. These 2 phases demand changes in the corporate culture, management of strategies and business model, as well as internal and external operations and relationships of the company.

At a SC level, collaboration is fundamental in the implementation of I4.0 technologies that have vertical and horizontal integration at their core [17]. Internal social capital has a high impact on the decision-making process of I4.0 technology adoption among the different adopter levels: "beginners", "adopters " and "non-adopters", whereas external social capital has a relevant impact only on "beginners" and "non-adopters" [1]. This is in line with the vision that the ability to effectively work in teams and leverage social contexts can contribute to the successful adoption of I4.0 technologies. To have a complete digital automation of the manufacturing process, covering both vertical and horizontal dimensions, employees are required to have an overall understanding of the organizational processes and information flows [12].

### 3.3.2.5. Management Support

Organizational and managerial practices in the SC and company processes directly affect the adoption of I4.0 technologies. "Management Support" acts as a moderating factor in the relationship between investment, social capital, and adoption, where social capital acts as a catalyst in the adoption process [1]. According to the aforementioned 2-phase framework, where digitalization of the business and transformation of the business network are the desired outcomes, management support is required to facilitate this process through changes in the corporate culture, management of strategies and business model, and organization of internal and external operations and relationships [12]. To support adoption, managers should encourage employees to make sense of the benefits of taking on new responsibilities [15], support decision-making, facilitate the reduction of hierarchical tiers and increase employee autonomy [1].

### 3.3.2.6. Absorptive Capacity

"Absorptive Capacity" moderates the relationship between social capital and the adoption process [1]. When the absorptive capacity" is high in organizations, it directly affects the adoption of I4.0 technologies [23]. Moreover, the "Exploratory Absorptive Capacity", *i.e.* the ability of companies to profit from the external knowledge flows, enable companies inserted in a certain cluster to adopt disruptive technologies before other companies that do not have this capacity. Exploiting is seen as a company's ability to improve, expand, and use its existing routines, skills, and technologies to create something new based on transformed knowledge [23].

### 3.3.3.    Political-Market Factors

### 3.3.3.1. Activity Sector

The "Activity Sector" factor has an impacting role in the adoption of I4.0 technologies, where generally the service sector is more susceptible to adoption than the manufacturing sector [11]. The scale of the organization also plays a role, since the adoption of certain technologies among SMEs is slower than in larger organizations, due to their limited resources [29].

### 3.3.3.2. Market Demand

The "Market Demand" factor has a high impact on the adoption process since the perception of value held by end customers generates a positive demand spiral for new benefits that consequently become part of the standard service offered by the company. This, in turn, generates competitive pressure in the market. The more devices that generate data on the corporate network and connect themselves to cloud solutions, the greater the possibility of creating value to customers; hence developing a differentiating factor in the market [16].

### 3.3.3.3. Government Policies and Regulations

The "Government Policies and Regulations" factor works as a key driver in the adoption process because legal information systems are needed in order to support the development and expansion of IoT in logistics and in Supply Chain Management (SCM); thereby enhancing the security standards to regulate operations. Government, institutions and organizations must work together to promote and support technological initiatives and solutions [16].

### 3.3.3.4. Standards and Validations

The existence of market standards and the ability for solutions from multiple vendors to work interchangeably can facilitate the adoption decision [11]. Validations play an important role in the adoption process [16], due to the scarcity of research on multiple applications of I4.0 technologies in industries. Due to the lack of standards, very few IoT technologies show clear returns on investment across the industry, which discourages SMEs in the adoption of innovative and disruptive technology [28].

### 3.3.4.    Technological Factors

### 3.3.4.1. IT Infrastructure

"IT Infrastructure" is directly linked with the quality and data generation in the SC. It is therefore one of the main factors impacting the decision-making process of I4.0 technology adoption. Poor IT infrastructure and internet connectivity prove to be substantial barriers to digital transformation or adoption of these technologies [12]. The "IT Infrastructure" factor forms the core of I4.0 technologies use. The implementation of IoT solutions requires advanced skill sets and infrastructural support [16], which are scarce and proves to be a critical challenge for adoption [20].

### 3.3.4.2. Systems Architecture, Integration and Compatibility

The "Systems Architecture, Integration and Compatibility" factor influences the adoption of I4.0 technologies, hence its volatility requires special attention from adopters [18]. Integration and compatibility issues have a significant impact on the adoption process of IoT technologies [16], as challenges in integrating IoT with existing legacy systems act as barriers to adoption [5].

### 3.3.4.3. Research and Development of Technologies & Data Management Quality

The "Research and Development" and "Data Management Quality" factors show significant impact on adoption since insufficient research and development practices in I4.0, absence of IT infrastructure, low-quality data, lack of digital culture, and distrust from partners create barriers to those organizations trying to innovate [12].

## 3.4. Benefits

The benefits of I4.0 technologies adoption herein presented are derived from the articles comprised in the SLR. Figure 5 displays the different technologies against their respective frequency of appearance in the 10 articles considered for this study.

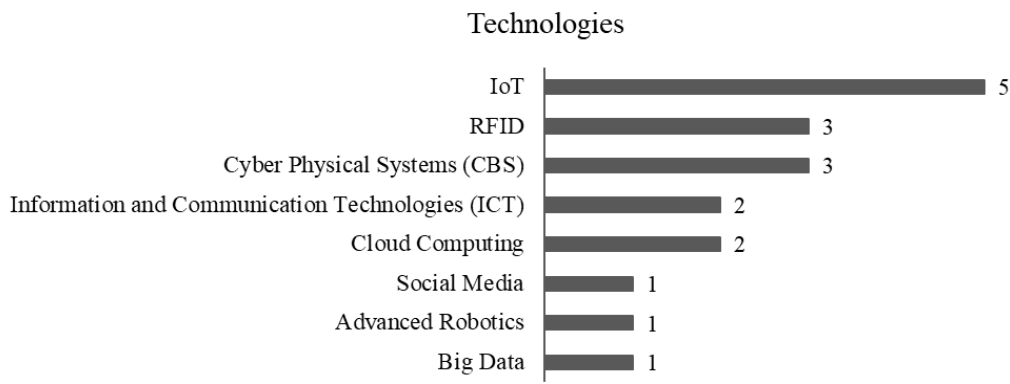

Figure 5. Frequency of technologies in studies

In the context of adapted SC, the adoption of radio-frequency identification (RFID) and cloud computing, which enable greater visibility and agility throughout the SC, leads to cost reductions, stabilization of inventory levels and increase of order fulfillment. Due to efficiency gains, the adoption benefits associated with I4.0 technologies in the SC are irrefutable, brought about by cyber-physical systems, RFID, IoT technologies, cloud computing, big data analytics, and advanced robotics [12].

Through absorptive capacity, innovation becomes a joint action between members, where the different relationships between organizations promote not only trust and other shared norms and values, but also the transmission of tacit knowledge [23]. Moreover, the adoption of IoT in operations and in the SC offers commercial benefits, including improved operating processes, low risks and costs, and increased productivity. IoT facilitates the search for new organizational capabilities, from a management and control perspective [16]. The data gathered from the IoT systems provide decision-makers with new ideas about value proposition and value creation, thereby helping to strengthen the bond with customers and to adopt more efficient policies and effective business practices [27]. Through enhanced visibility, transparency, adaptation, flexibility and virtualization in SC [32], companies with the greatest innovation capacity experience increased global competitiveness, reduction in costs, increased market share and increased quality of services [11]. Figure 6 illustrates the benefits and their respective frequency of appearance in the SLR.

Benefits in the SC

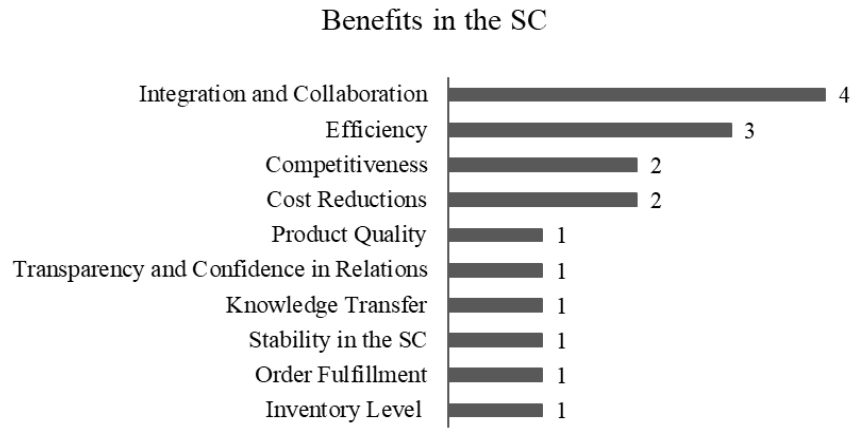

Figure 6. Frequency of benefits in studies

## 3.5. Results

The Framework below (Figure 7) displays the relationship, in a systematic fashion, between adoption factors, I4.0 technologies and the benefits yielded from adoption in the SC.

Factors such as "Perception of Use", "Perception of Ease of Use" and "Acquisition and Operating Costs" in AMTs directly affect the intention to adopt I4.0 technologies. Both "Internal and External Social Capital" are mediated by the "Management Support" and "Absorptive Capacity". Additionally, "Systems Architecture", "Standards and Validations", "Integration and Compatibility", and "Security and Privacy" are highly volatile and have an impact on one other, which may affect the adoption of technology. "Government Policies and Regulations" together with "IT Infrastructure" are determining factors of this adoption structure, forming the basis of integration of processes and performance.

The benefits of I4.0 technology adoption in organizations lead us to reduced operating costs through increases in efficiency in the manufacturing and logistics processes. With a high level of process and IT integration in the SC, organizations can improve their market competitiveness. Finally, they can foster innovation along the SC due to collaboration and knowledge transfer.

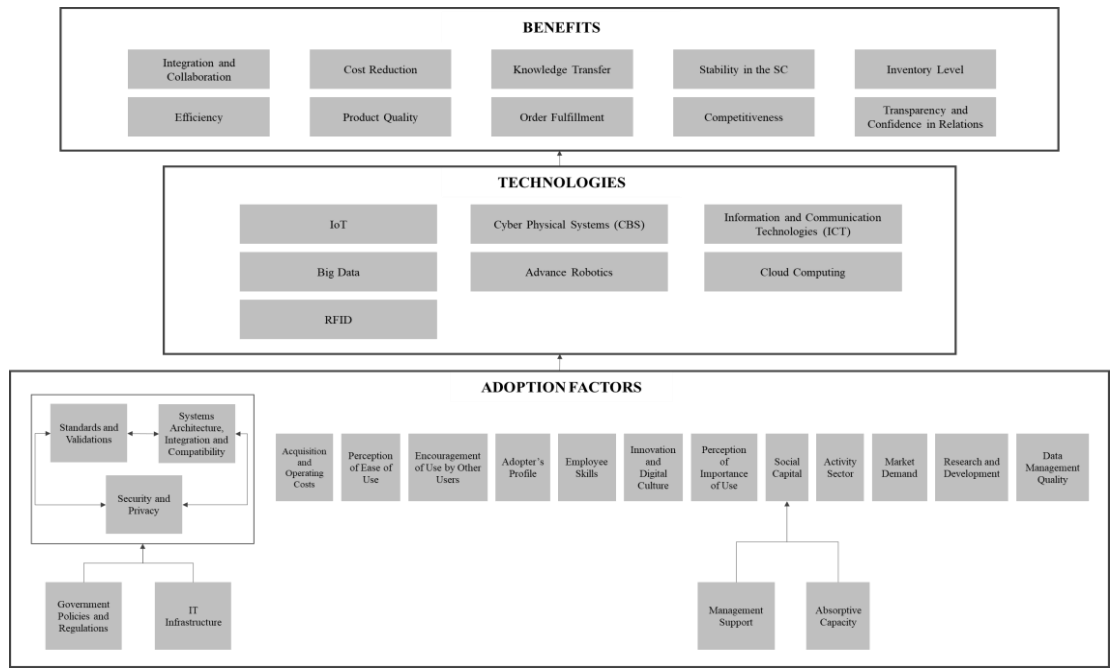

Figure 7. Adoption framework

## 4. CONCLUSION

The SLR demonstrated, through the results of relevant studies, that the existence or absence of impacting factors, as well as their intensity, affects the adoption of I4.0 technologies. As proposed in the objective of this work, the results from this study contribute to the academic literature related to I4.0 technology adoption by presenting a framework that integrates adoption factors, I4.0 technologies and benefits to the SC. The framework proposed in this study can be easily adapted to serve as a tool in the assessment and selection of technological innovations. This study paves the way for organizations in the adoption process of I4.0 technologies to understand the challenges related to the adoption factors and introduce the potential benefits that can add value to the SC, thereby guiding entrepreneurs in their digital transformation journey.

## 5. LIMITATIONS AND FURTHER STUDIES

This study has the following limitations:

First: Despite the study showcasing a broad scope of technologies; several other enabling I4.0 technologies did not take part in this study, *e.g.* Extended Reality, 3D Printing and Simulations. Further studies should also try to encompass these technologies.

Second: Due to the broad approach of this study, the adoption process for specific technologies was not deeply explored. Further studies should steer the focus toward a specific technology.

## AUTHORS

*José Carlos Franceli*

Student of the Master's Program in Production Engineering at the Federal University of ABC (UFABC) with thesis theme "Role of the Top Management in the IoT adoptions in OEM Suppliers and the Supply Chain benefits". Professional in the Automotive Industry with more than 30 years experience in Supply Chain Management in Brazil, USA and Germany.

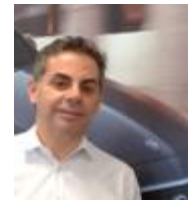

*Silvia Novaes Zilber Turri*

Adjunct professor of the Undergraduate course in Management Engineering and Coordinator of the Master's Program in Production Engineering at the Federal University of ABC (UFABC). Collaborator professor at FEA / University of São Paulo and a full professor at the Postgraduate Program in Management (PPGA) at Universidade 9 de Julho. Evaluator of national and international journals. With published articles in several international journals in the area of Innovation and Digital Transformation of the Value Chain, Business Models for Innovation and Innovation in Small Companies.

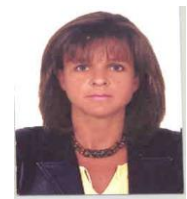